%% file: desy05-065.tex
\documentclass[12pt]{article}
\usepackage{epsfig}
\usepackage{lineno}
\usepackage{amsmath}
\usepackage{hhline}
\usepackage{amssymb}
\usepackage{times}
\usepackage{cite}
\usepackage{multirow}
\usepackage{array}

\newlength{\dinwidth}
\newlength{\dinmargin}
\begin{document}  

\newcommand{\pom}{{I\!\!P}}
\newcommand{\reg}{{I\!\!R}}
\newcommand{\gap}{\stackrel{>}{\sim}}
\newcommand{\lap}{\stackrel{<}{\sim}}

\newcommand{\alps}{\alpha_s}
\newcommand{\sqrts}{$\sqrt{s}$}
\newcommand{\LO}{$O(\alpha_s^0)$}
\newcommand{\Oa}{$O(\alpha_s)$}
\newcommand{\Oaa}{$O(\alpha_s^2)$}
\newcommand{\PT}{p_{\perp}}
\newcommand{\PO}{I\!\!P}
\newcommand{\xpomlo}{3\times10^{-4}}  
\newcommand{\dgr}{^\circ}
\newcommand{\pbarnt}{\,\mbox{{\rm pb$^{-1}$}}}

\def\lfrestriction#1{\lower.25ex\hbox{\Big|}_{#1}} 
%
%
\newcommand{\tg}{\theta_{\gamma}}
\newcommand{\te}{\theta_e}
%
%
\newcommand{\qsq}{\mbox{$Q^{\,2}$}}
\newcommand{\Qsq}{\mbox{$Q^2$}}
\newcommand{\s}{\mbox{$s$}}
\newcommand{\ttra}{\mbox{$t$}}
\newcommand{\modt}{\mbox{$|t|$}}
\newcommand{\eminpz}{\mbox{$E-p_z$}}
\newcommand{\eminpzs}{\mbox{$\Sigma(E-p_z)$}}
\newcommand{\rap}{\ensuremath{\eta^*} }
\newcommand{\W}{\mbox{$W$}}
\newcommand{\w}{\mbox{$W$}}
\newcommand{\Q}{\mbox{$Q$}}
\newcommand{\q}{\mbox{$Q$}}
\newcommand{\xB}{\mbox{$x$}}  
\newcommand{\xF}{\mbox{$x_F$}}  
\newcommand{\xg}{\mbox{$x_g$}}  
\newcommand{\xbj}{x}
\newcommand{\xpom}{x_{\PO}}
\newcommand{\y}{\mbox{$y~$}}

%
\newcommand{\gp}{\ensuremath{\gamma^*}p }
\newcommand{\gammasp}{\ensuremath{\gamma}*p }
\newcommand{\gammap}{\ensuremath{\gamma}p }
\newcommand{\gsp}{\ensuremath{\gamma^*}p }
\newcommand{\epem}{\mbox{$e^+e^-$}}
\newcommand{\ep}{\mbox{$ep~$}}
\newcommand{\epl}{\mbox{$e^{+}$}}
\newcommand{\emi}{\mbox{$e^{-}$}}
\newcommand{\epm}{\mbox{$e^{\pm}$}}
%
\newcommand{\photon}{\mbox{$\gamma$}}
\newcommand{\phib}{\mbox{$\varphi$}}
\newcommand{\rh}{\mbox{$\rho$}}
\newcommand{\rhz}{\mbox{$\rh^0$}}
\newcommand{\ph}{\mbox{$\phi$}}
\newcommand{\om}{\mbox{$\omega$}}
\newcommand{\ome}{\mbox{$\omega$}}
\newcommand{\jpsi}{\mbox{$J/\psi$}}
\newcommand{\JPSI}{J/\psi}
\newcommand{\ups}{\mbox{$\Upsilon$}}
\newcommand{\bsl}{\mbox{$b$}}
%
%
\newcommand{\cm}{\mbox{\rm cm}}
\newcommand{\GeV}{\mbox{\rm GeV}}
\newcommand{\gev}{\mbox{\rm GeV}}
\newcommand{\GeVx}{\rm GeV}
\newcommand{\gevx}{\rm GeV}
\newcommand{\MeV}{\mbox{\rm MeV}}
\newcommand{\mev}{\mbox{\rm MeV}}
\newcommand{\MeVx}{\mbox{\rm MeV}}
\newcommand{\mevx}{\mbox{\rm MeV}}
\newcommand{\GeVsq}{\mbox{${\rm GeV}^2$}}
\newcommand{\gevsq}{\mbox{${\rm GeV}^2$}}
\newcommand{\gevsqc}{\mbox{${\rm GeV^2/c^4}$}}
\newcommand{\gevcsq}{\mbox{${\rm GeV/c^2}$}}
\newcommand{\mevcsq}{\mbox{${\rm MeV/c^2}$}}
\newcommand{\GeVsqm}{\mbox{${\rm GeV}^{-2}$}}
\newcommand{\gevsqm}{\mbox{${\rm GeV}^{-2}$}}
\newcommand{\nb}{\mbox{${\rm nb}$}}
\newcommand{\nbinv}{\mbox{${\rm nb^{-1}}$}}
\newcommand{\pbinv}{\mbox{${\rm pb^{-1}}$}}
\newcommand{\mm}{\mbox{$\cdot 10^{-2}$}}
\newcommand{\mmm}{\mbox{$\cdot 10^{-3}$}}
\newcommand{\mmmm}{\mbox{$\cdot 10^{-4}$}}
\newcommand{\degr}{\mbox{$^{\circ}$}}
%
%
\def\gsim{\,\lower.25ex\hbox{$\scriptstyle\sim$}\kern-1.30ex%
  \raise 0.55ex\hbox{$\scriptstyle >$}\,}
\def\lsim{\,\lower.25ex\hbox{$\scriptstyle\sim$}\kern-1.30ex%
  \raise 0.55ex\hbox{$\scriptstyle <$}\,}
\begin{titlepage}

\noindent

\begin{flushleft}
DESY 05-065 \hfill ISSN 0418-9833 \\
April 2005
\end{flushleft}

\vspace{2cm}

\begin{center}
\begin{Large}

{\boldmath \bf
Measurement of Deeply Virtual Compton Scattering\\
 at HERA }
\vspace{2cm}

H1 Collaboration

\end{Large}
\end{center}

\vspace{2cm}

\begin{abstract}

A measurement is presented of elastic deeply virtual Compton 
scattering $\gamma^* p \rightarrow \photon p$ made using  $e^+ p$ collision data 
corresponding to a luminosity of 46.5 pb$^{-1}$, taken with the H1 detector at HERA.
The cross section is measured as a function of the photon 
virtuality, $Q^{\,2}$, the invariant mass of the  
\gp\ system, $W$, and for the first time, differentially in the 
squared momentum transfer at the proton vertex, $t$, in the kinematic range 
$2~<~Q^{\,2}~<~80\,{\rm GeV}^2$,  $30~<~W~<~140\,{\rm GeV}$ and 
$|t|~<~1\,{\rm GeV}^2$.  QCD based calculations at next-to-leading order 
using generalized parton distributions can describe the data,
as can colour dipole model predictions.

\end{abstract}

\vspace{1.5cm}

\begin{center}
To be submitted to Eur. Phys. J. C.
\end{center}

\end{titlepage}

\begin{flushleft}
  \input{h1auts}
\end{flushleft}

\newpage
\section{Introduction}
\noindent


Measurements of the deep-inelastic scattering (DIS) of leptons and nucleons
allow the extraction of Parton Distribution Functions (PDFs) which describe
the longitudinal momentum carried by the quarks, anti-quarks and gluons that
make up the fast-moving nucleons. While these PDFs provide crucial input to
perturbative Quantum Chromodynamic (QCD) calculations of processes involving
hadrons, they do not provide a complete picture of the partonic structure of
nucleons. In particular, PDFs contain neither information on the
correlations between partons nor on their transverse motion. This missing
information can be provided by measurements of processes in which the
nucleon remains intact, such as the exclusive production of light meson
states in lepton-nucleon collisions, and is encoded in Generalised Parton
Distributions (GPDs)~\cite{Muller:1994fv, Ji:1997ek,Radyushkin:1997ki,diehlreport}.

\begin{figure}[htb]
 \begin{center}
  \epsfig{figure=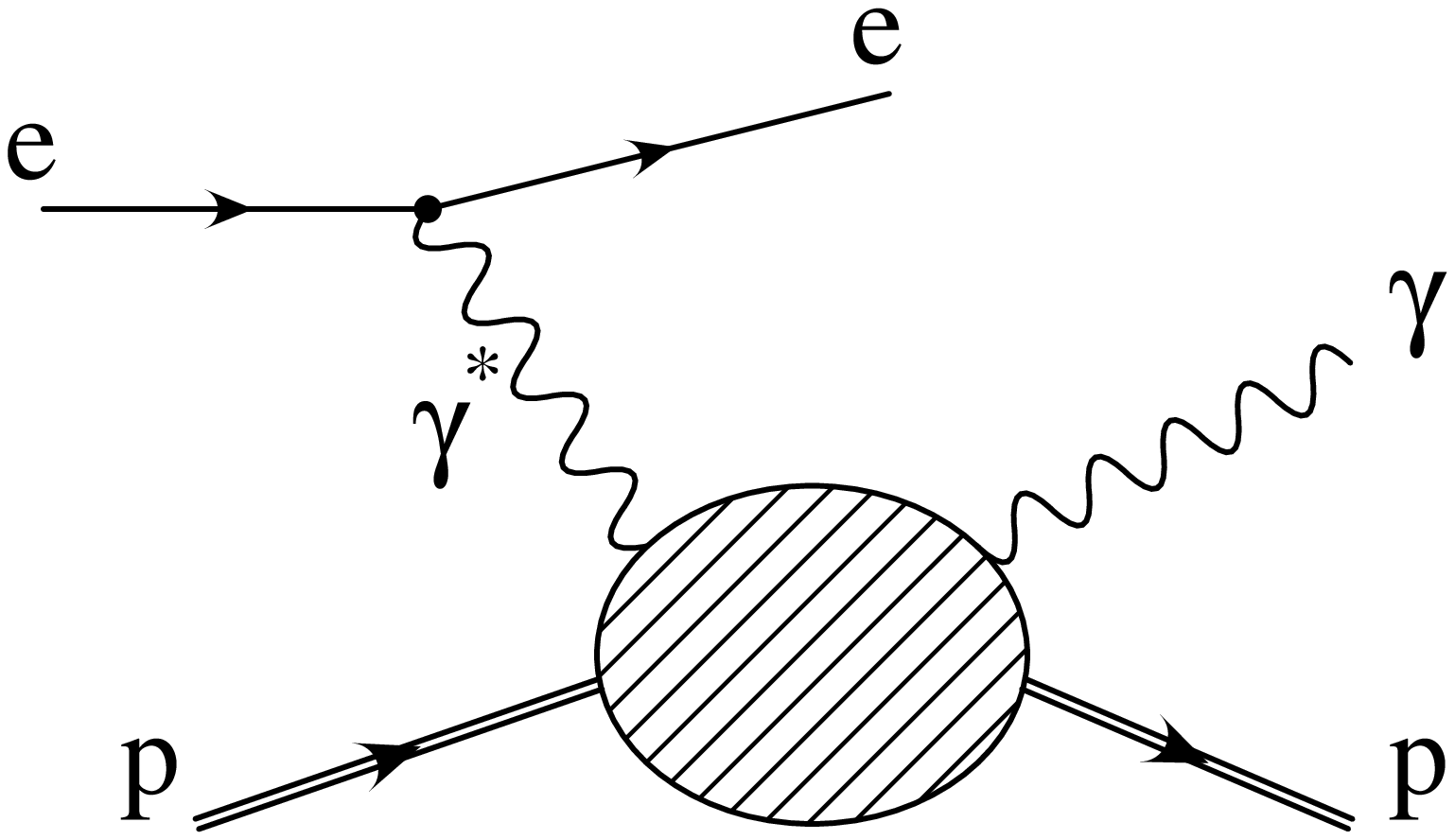,height=0.22\textwidth}
  $\qquad$
  \epsfig{figure=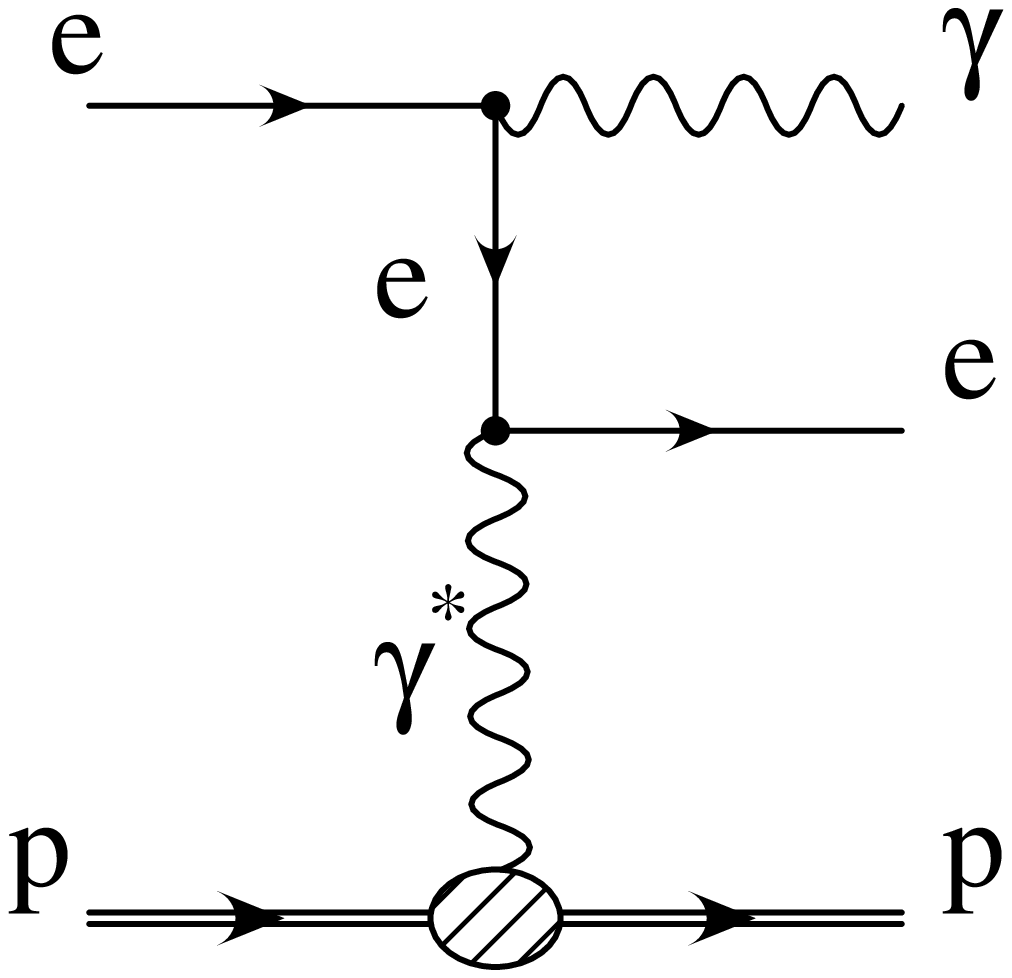,height=0.22\textwidth}
  $\qquad$
  \epsfig{figure=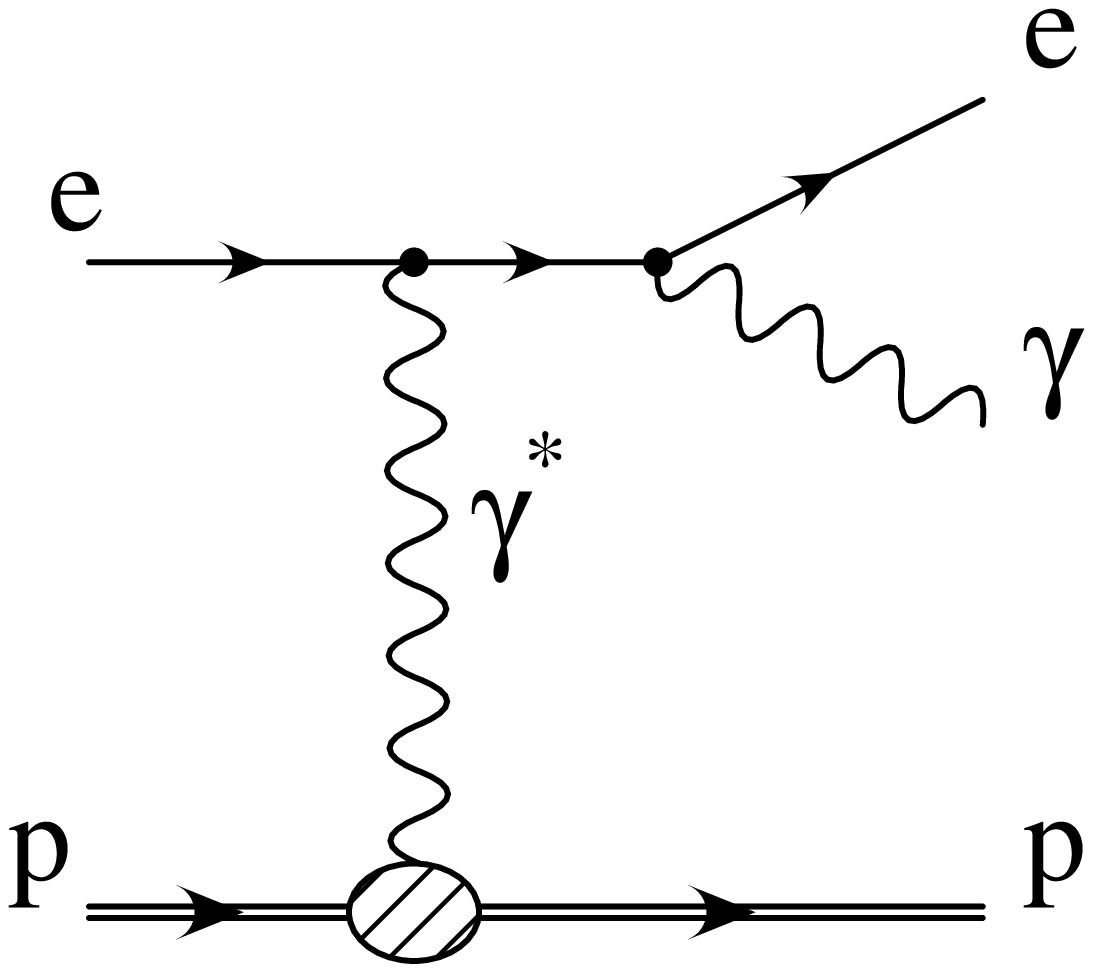,height=0.22\textwidth}
  \\
  \begin{picture}(0,0)(100,0)
  \put(52,-1){\bf a)}
  \put(107,-1){\bf b)}
  \put(150,-1){\bf c)}
  \end{picture}
  \caption{\sl Diagrams illustrating the DVCS (a) and the Bethe-Heitler 
    (b and c) processes.}
  \label{fig:bh}
 \end{center}
\end{figure}

The simplest process sensitive to GPDs is deeply virtual
Compton scattering (DVCS) (figure~\ref{fig:bh}a), which is the diffractive
scattering of a virtual photon off a proton~\cite{Ji:1997nm, Collins:1999be, 
Ji:1998xh, Mankiewicz:1998bk, Belitsky:2000sg, Blumlein:2000cx},
$\gamma^* p \rightarrow \gamma p$. In the present analysis
DVCS is accessed through the reaction:

\begin{equation}
e^+  p \rightarrow e^+  \photon  p .
\end{equation}
\label{eq:reac}

This process is of particular interest as it has both a clear
experimental signature and is calculable in perturbative QCD: 
it does not suffer from the uncertainties caused
by the lack of understanding of the meson wave function that plague
exclusive vector meson electroproduction.

The reaction studied receives contributions from both the DVCS 
process, whose origin lies in the strong interaction, and the purely 
electromagnetic Bethe-Heitler (BH) process (figures~\ref{fig:bh}b and 
\ref{fig:bh}c), where the photon is emitted from the positron. 
The BH cross section can be precisely calculated  in QED 
using elastic proton form factors.
Here, the DVCS cross section is obtained by subtracting the BH
contribution from the total cross section, which is possible 
since the interference contribution
vanishes~\cite{Belitsky:2000sg}, as this measurement is integrated over azimuthal angles. 

The first measurements of the DVCS cross section at high energy were
obtained by H1~\cite{h1dvcs97} and ZEUS~\cite{zeusdvcs} 
and the helicity asymmetry in DVCS has been measured at lower energy 
with polarised lepton beams by HERMES~\cite{hermesdvcs} 
and CLAS~\cite{clasdvcs}.

In this paper, a measurement of the DVCS cross section is presented,
based on data collected with the H1 detector at HERA in the years 1996
to 2000. 
These data correspond to a luminosity of 46.5 \pbinv, a factor
of 4 larger than the luminosity used in the previous H1
publication~\cite{h1dvcs97}, which is based only on 1997 data.
The cross section is presented as a function of the photon 
virtuality, $Q^2$, the invariant mass of the  
\gp\ system, $W$, and the  squared momentum transfer at the proton vertex, $t$.

\section{Generalized Parton Distributions and Theoretical Predictions}

\begin{figure}[htb]
 \begin{center}
   \epsfig{figure=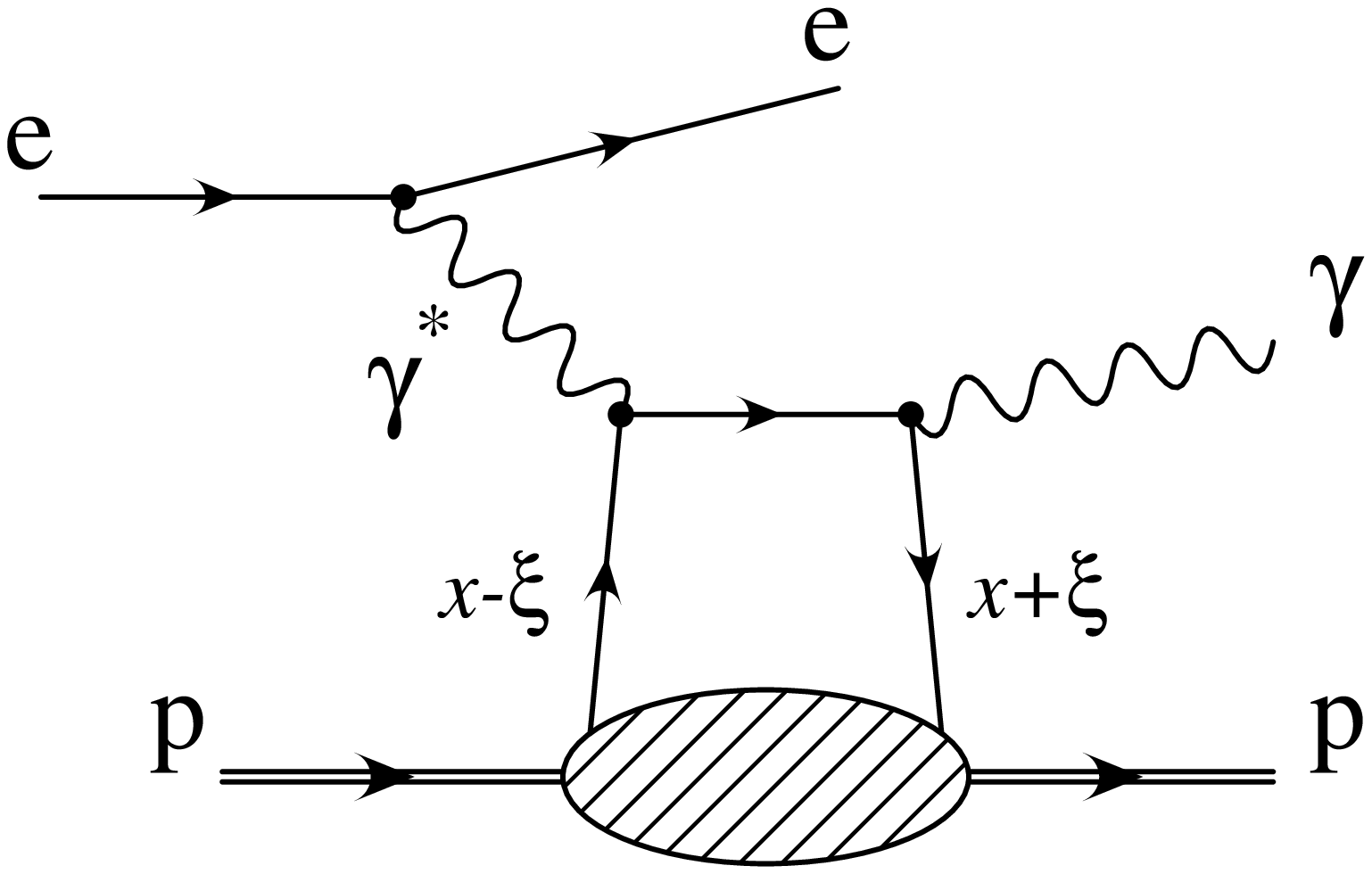,width=0.40\textwidth}
  $\qquad$
  \epsfig{figure=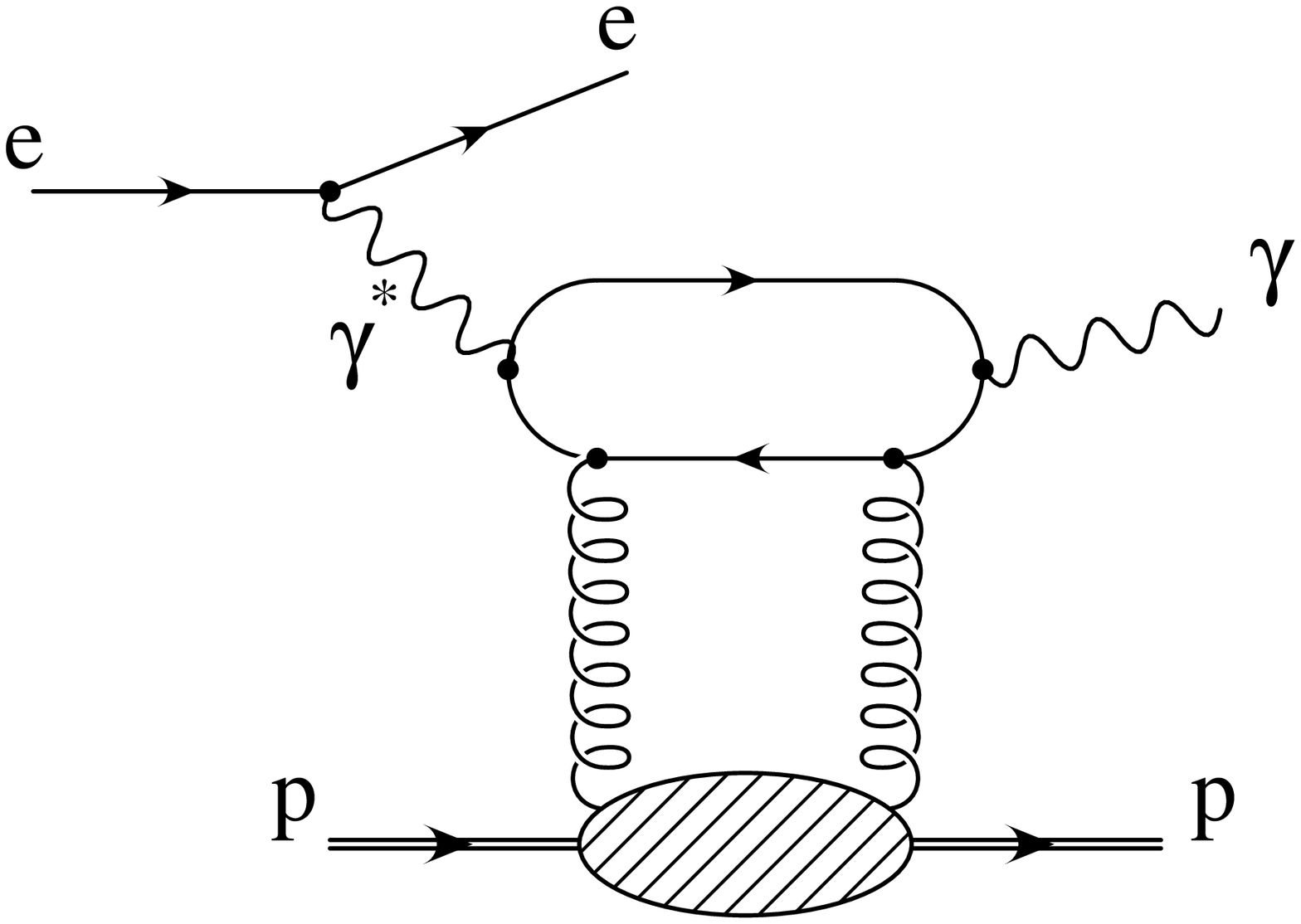,width=0.40\textwidth}
  \\
  \begin{picture}(0,0)(100,0)
  \put(65,-1){\bf a)}
  \put(140,-1){\bf b)}
  \end{picture}
  \caption{\sl Examples of diagrams for the DVCS process a) at leading
        order, b) at next-to-leading order.}
  \label{fig:dvcs}
 \end{center}
\end{figure}

The leading order diagram for DVCS in positron proton scattering is 
shown in figure~\ref{fig:dvcs}a and 
a diagram that contributes at next-to-leading order in figure~\ref{fig:dvcs}b.
The transition from a virtual photon to a real photon
forces the fractional momenta of the two partons involved to be different (``skewed").
Hence, DVCS is sensitive to the correlations between partons in the proton which are
encoded in the GPDs. In the presence of a hard scale, here $Q^2$, the 
DVCS scattering amplitude factorises~\cite{Radyushkin:1997ki,Collins:1999be,Ji:1998xh}
into a hard part, calculable order by order in perturbative QCD,
and the GPDs which contain the non-perturbative 
effects due to the structure of the proton.

\subsection{Generalized Parton Distributions}

The GPDs generalize and interpolate between the PDFs and elastic form factors.
The PDFs  contain information on the longitudinal momenta 
of the partons while form factors contain information on their transverse momenta, 
often in the form of sum rules related to 
charges, local currents and the energy-momentum tensor of QCD.
GPDs have simple physical significance in light-cone
coordinates (or the infinite momentum frame), where
they represent the interference of two different wave functions, one
of a parton having a momentum fraction $x + \xi$ and the other of a parton
 with a momentum fraction  $x - \xi$, as is illustrated in figure~\ref{fig:dvcs}. 
Besides the longitudinal momentum fraction
variables $\xi$ (called skewedness) and $x$, GPDs depend on $t$, 
the square of the four-momentum exchanged at the hadron vertex.
GPDs are defined at a starting scale $\mu^2$ and their $Q^2$ evolution is
generated by perturbative QCD.

 There are two different types of GPDs (for a quark $q$ or a
gluon $g$) in the unpolarised case: $H^{q,g}(x, \xi, t)$ and $ E^{q,g}(x, \xi, t)$.  
While the $E^{q,g}$ distributions have no equivalent in the ordinary PDF approach,
the $H^{q,g}$ reduce to the usual PDFs in the forward limit $(\xi~=~0, \; t~=~0)$, 
{\it i.e.}~$ H^q(x,0,0) = q(x)$ and $ H^g(x,0,0) = x g(x)$,
where $q(x)$ and $g(x)$ are the ordinary parton distributions. 
The variable $x$ is defined in the range $[-1,1]$, with negative values
corresponding to anti-quark distributions: $H^q(-x,0,0)=-\bar{q}(x)$.
The gluon GPD is symmetric in $x$ in the forward limit: $H^g(-x,0,0)=H^g(x,0,0)$.
The skewedness variable $\xi$ is related to the well known Bjorken-$x$
variable, $x_{Bj}$, by 
$\xi = \alpha x_{Bj} / ( 2- \alpha x_{Bj} ) $,
where $\alpha = 1 + q'^2/Q^2$ and $q'$ denotes the four-momentum of the outgoing 
photon\footnote{
		For the DVCS process, the outgoing photon is real ($q'^2=0$)
		and $\xi$ reduces to $x_{Bj}/(2-x_{Bj})$. The forward limit
		corresponds to the case of inclusive DIS, where $q'^2 = -Q^2$
		and thus $\xi = 0$.
               }.
The first moments of the GPDs in $x$ are given by form factors~\cite{diehlreport}.

Two different kinematic regions exist for GPDs with respect to the variables $x$
and $\xi$. The DGLAP region, where $|x|>\xi$~\cite{Gribov:1972ri, 
Lipatov:1974qm, Altarelli:1977zs, Dokshitzer:1977sg},
corresponds to the emission and re-absorption of a quark, 
anti-quark or a gluon. The ERBL~\cite{Lepage:1979zb, Efremov:1979qk} region, 
where $|x| < \xi$, corresponds to meson or gluon pair exchange. 
Each region has its own evolution equations.

The recent strong interest in GPDs was stimulated by the information
they contain on the spin structure of the nucleon. 
In particular, GPDs are so far the only known means of probing 
the orbital motion of partons in the nucleon through
Ji's Sum Rule \cite{ref:spin},
which relates unpolarised GPDs to the total angular momentum of the proton.
DVCS measurements at HERA can provide constraints on this
sum rule through their sensitivity to the GPDs.

\subsection{Theoretical Predictions} \label{sect:freund}

The measurements presented here are compared with NLO QCD calculations
and predictions made using colour dipole approaches.
In NLO QCD, the DVCS cross section has been calculated~\cite{Freund:2001hm,Freund:2001hd} 
using two different GPD parameterisations~\cite{Freund:2002qf}.
The  $t$ dependence of the GPDs is taken to be $e^{-b|t|}$. 
The MRST2001~\cite{MRS2001} and CTEQ6~\cite{CTEQ6.1} parameterisations of the PDFs
are used in the DGLAP region ($|x|>\xi$). Thus $H$, which provides the main contribution 
to DVCS at small $x_{Bj}$, is given at the starting scale $\mu$ by
$H^q(x,\xi,t)=q(x) \, e^{-b|t|}$ for the quarks
and $H^g(x,\xi,t)=x\ g(x) \, e^{-b|t|}$ for the gluons\footnote{A different ansatz 
for GPDs has been used in~\cite{Belitsky:2001ns} in a LO calculation of the DVCS cross section.}. 
Both the skewing and the $Q^{2}$ dependence are generated dynamically.
In the ERBL region ($|x| < \xi$), these parameterisations 
have to be modified, ensuring a smooth continuation to the DGLAP region
(for details see~\cite{Freund:2002qf}).
These GPD models are found to describe both the shape of the previous
H1 DVCS cross section measurements~\cite{h1dvcs97} and the single spin asymmetry 
measured by HERMES~\cite{hermesdvcs}.

The DVCS cross section has also been calculated in the colour dipole
approach, which is successful in describing both inclusive and
diffractive scattering in the DIS regime at high energy.
These predictions are based on a factorisation of the DVCS amplitude into the wave function
for the photon to fluctuate into a $q \bar q$ pair, the cross section for this pair
to interact with the proton and the outgoing photon wave function.  
If $s$-channel helicity is conserved in DVCS, the virtual photon must be transversely
polarised. As the wave function of the transversely polarised $\gamma^*$ can select
large dipole sizes, whose interactions are predominantly soft, DVCS constitutes a good probe
of the transition between the perturbative and non-perturbative regimes of QCD.
The various calculations differ in the way the dipole cross section is parameterised.
Donnachie and Dosch~\cite{Donnachie:2000px}
use soft and hard pomeron exchange depending on the size of the dipole.
All parameters are determined from $pp$ and $\gamma^*p$ total cross section
measurements. Favart and Machado~\cite{Favart:2003cu} apply the saturation model
of Golec-Biernat {\it et al.}~\cite{Golec-Biernat:1999qd} to the DVCS process
and use DGLAP evolution~\cite{Favart:2004uv},
following the approach of Bartels, Golec-Biernat and Kowalski (BGBK)~\cite{Bartels:2002es}.
In both cases an exponential $t$-dependence, $e^{-b|t|}$, is assumed.

\section{Experimental Procedure}

\subsection{H1 Detector}

A detailed description of the H1 detector can be found in~\cite{h1dect}.
Here only the detector components relevant for the present analysis are
described. 
The SpaCal~\cite{spacal}, a lead scintillating fibre calorimeter, 
covers the backward\footnote{H1 uses a right-handed coordinate system with $z$ axis along
the beam direction, the $+z$ or ``forward" direction being that of the outgoing proton beam.
The polar angle $\theta$ is defined with respect to the $z$ axis and the
pseudo-rapidity is given by $\eta=-\ln \tan \theta /2$.}
region of the H1 detector ($ 153 ^{\rm \circ} < \theta < 177.5 ^{\rm \circ}$).
Its energy resolution for electromagnetic showers is $\sigma(E)/E
\simeq 7.1\%/\sqrt{E/{\rm GeV}} \oplus 1\%$. 
The liquid argon (LAr) calorimeter ($4^{\rm \circ} \leq \theta \leq
154^{\rm \circ}$) is situated inside a solenoidal magnet. 
The energy resolution for electromagnetic showers is 
$\sigma(E)/E \simeq 11\%/\sqrt{E/{\rm GeV}}$ as obtained from test beam 
measurements~\cite{Andrieu:1994yn}.
The backward drift chamber (BDC), placed in front of the SpaCal,
measures track segments for charged particles entering the SpaCal from the
interaction region. These are used to identify the scattered
positron and to determine its position with a resolution of 0.5~mm in the
radial and 2.5~mm in the azimuthal direction.
The main component of the central tracking detector is the central jet
chamber (CJC) which consists of 
two \mbox{2\,m} long coaxial cylindrical drift chambers, 
with wires parallel to the beam direction.
The measurement of charged particle transverse momenta is performed
in a magnetic field of 1.15~T, uniform over the full tracker volume.
The forward components of the detector, used here to tag hadronic
activity at large pseudo-rapidity ($5 \lsim \eta \lsim 7$), are
the forward muon detector (FMD) and the proton remnant tagger (PRT).
The FMD, designed to identify muons 
emitted in the forward direction, contains six planes of drift cells.
It is used here to detect the particles produced when a proton dissociates and
secondary interactions occur in the beampipe and adjacent material.
Secondary particles, or the scattered proton, can 
also be detected by the PRT, which is located at 24 m from the interaction 
point and consists of layers of scintillator surrounding 
the beam pipe. 
The luminosity is determined from the rate of BH events measured in a
luminosity monitor.

\subsection{Kinematics}

For DVCS, the final state photon
does not originate from the positron and therefore the ratio of the DVCS to the BH cross 
sections is expected to increase when the photon is scattered in the forward direction.
The analysis sample is thus selected by requiring a photon candidate in the LAr calorimeter
and a positron candidate in the SpaCal calorimeter.

The reconstruction of the kinematic variables $Q^2$, 
$x_{Bj}$ and $W$ relies on the polar angle measurements of the final 
state positron, $\te$, and photon, $\tg$:

\begin{eqnarray}
Q^2 & = & 4 E_0^2\, \frac{\sin\ \tg\ (1+\cos \te)}
                {\sin\ \tg\ + \sin\ \te\ - \sin\ (\te\ + \tg)} \; , \\
x_{Bj} & = & \frac{E_0}{E_p} \, \frac{\sin\ \tg\ + \sin\ \te\ + \sin\ (\te\ +
\tg)}
                {\sin\ \tg\ + \sin\ \te\ - \sin\ (\te\ + \tg)} \;
        \quad \rm{and} \\
   W^2 & =&  \frac{Q^2}{x_{Bj}}\, (1-x_{Bj}) \; ,
\end{eqnarray}
where $E_0$ and $E_p$ are the positron and proton beam
energies, respectively.
For the majority of the events, the scattered positron trajectory is not
measured in the CJC and the event vertex cannot be determined.
The polar angles of the positron and photon are then determined assuming that 
they come from the nominal event vertex.
The square of the four-momentum transfer to the
proton, $t$, is very well approximated by the square of the 
vector sum of the transverse momenta 
of the final state photon, $\vec p_{t_{\gamma}}$, and of the scattered 
positron, $\vec p_{t_{e}}$:
 
\begin{equation}
  t \simeq - (\vec p_{t_{\gamma}} + \vec p_{t_{e}})^2 \ .
\label{eq:t}
\end{equation}

\subsection{Monte Carlo Simulation}  \label{sect:mc}

Monte Carlo (MC) simulations are used to estimate the corrections that must be
applied to the data due to the finite acceptance and resolution of the  detector.
Elastic DVCS events in $ep$ collisions are generated using the Monte Carlo generator
MILOU~\cite{Milou}, which is based on a NLO QCD cross section 
calculation~\cite{Freund:2001hm,Freund:2001hd,Freund:2003qs} (see
section~\ref{sect:freund}), and using a slope in $t$ of $b=6$~GeV$^{-2}$.
Higher order photon radiation from the incoming positron is 
implemented in the collinear approximation.
DVCS events in which the proton dissociates into a baryonic system $Y$ are
also simulated with the program MILOU using a $t$ slope of
$b_{pdiss}=1.5$~GeV$^{-2}$~\cite{Xavier}.
The Monte Carlo generator COMPTON~2.1~\cite{compton2, compton21} is used to 
simulate both elastic and inelastic BH events. 
Hadronisation
processes in inelastic BH events are simulated using the
SOPHIA model~\cite{Mucke:2000yb}.
Diffractive \om\ and \ph\ meson events are generated with the 
DIFFVM Monte Carlo program~\cite{diffvm}.
The events generated using all these programs are passed through a detailed simulation of the H1
detector and are subject to the same reconstruction and analysis chain
as the data.

\subsection{Event Selection}  \label{sect:selection}

The data were obtained with the H1 detector
when the HERA collider was operated with 820~\gev\ (1996-1997) and
920~\gev\ (1999-2000) protons and 27.6~\gev\
positron beams. The data sample corresponds to an integrated luminosity of 46.5~\pbinv,
 11.5~\pbinv\ of which were accumulated in 1996-1997 and 35~\pbinv\ in 1999-2000.
The event trigger used is based on the detection of an energy deposition 
greater than 6~GeV in the electromagnetic section of the SpaCal calorimeter. 
Due to the different trigger settings, 
selected events in the 1996-1997 period are in the kinematic range $Q^2~>~2$~GeV$^2$ 
while those in the 1999-2000 period are in the range $Q^2~>~4$~GeV$^2$. 

The DVCS event selection requires that the following criteria be
fulfilled.  The scattered positron must be detected in the SpaCal, have 
an energy larger than $15\,{\rm GeV}$ and be validated by a track segment in the BDC.
The photon must be measured in the LAr calorimeter with 
a transverse momentum $p_t > 1\,{\rm GeV}$ (1996-1997) or $p_t > 1.5\,{\rm GeV}$
(1999-2000) and a polar angle between $25^{\circ}$ and $145^\circ$. 
The scattered proton escapes undetected through the beam pipe.
Events with more than one central track are rejected while events with one central track 
are only kept if that track is associated with the scattered positron.
In order to reject inelastic and proton dissociation events, 
no further energy deposition in the LAr calorimeter with energy above 
$0.5\,{\rm GeV}$ is allowed and no activity above 
the noise level is allowed  in the PRT and FMD. 
The influence of QED radiative corrections is reduced by the requirement that
the longitudinal momentum balance $\sum (E - P_z) > 45\,{\rm GeV}$. 
Here, $E$ denotes the energy and $P_z$ the momentum along the beam axis of the 
final state particles and the sum runs over 
all such particles. To enhance the DVCS signal with respect to the BH 
contribution and to ensure a large acceptance, the 
kinematic domain is explicitly restricted to
$Q^2~<~80\,{\rm GeV}^2 $, $ |t|~<~1\,{\rm GeV}^2$ and
$30~<~W~<~140\,{\rm GeV}$. 


The selected sample contains 1243 events and is dominated
by the DVCS contribution, but  also contains contributions from the elastic
BH process and from the (inelastic) BH and DVCS processes with proton
dissociation, $ e^+  p \rightarrow e^+  \photon  Y$,
where the baryonic system $Y$ of mass $M_Y$ is not detected in the forward detectors.

As in previous H1 DVCS analyses \cite{h1dvcs97,Rainer}, a control
sample of BH events is also selected. Here, it is required that the
positron be detected in the LAr and the photon in the SpaCal.
It has been verified that the COMPTON MC correctly describes the
normalisation and the shapes of the distributions of the kinematic
variables for these events within an uncertainty of 5\%. 
Using events with a signal in the forward detectors, and
subtracting the inelastic BH contribution, obtained from the COMPTON MC,
the contribution of proton dissociation to the DVCS event sample 
is estimated to be $16\pm 8\,\%$ for the 1996-1997 data
(lower $Q^2$) and $10\pm 5\,\%$ for the 1999-2000 data.
The other backgrounds considered are diffractive 
\om\ and \ph\ production, with decay modes to final states including photons.
The main backgrounds originate from the decays
$\omega \rightarrow \pi^0 \gamma$ and $\phi \rightarrow K^0_L K^0_S$  
followed by the decay $K^0_S \rightarrow \pi^0 \pi^0$.
The contribution of these processes to the DVCS sample is estimated to be
below 3.5\% for the data taken in 1996-1997 and below 1\% for that taken in 1999-2000.

\begin{figure}
 \begin{center}
  \epsfig{figure=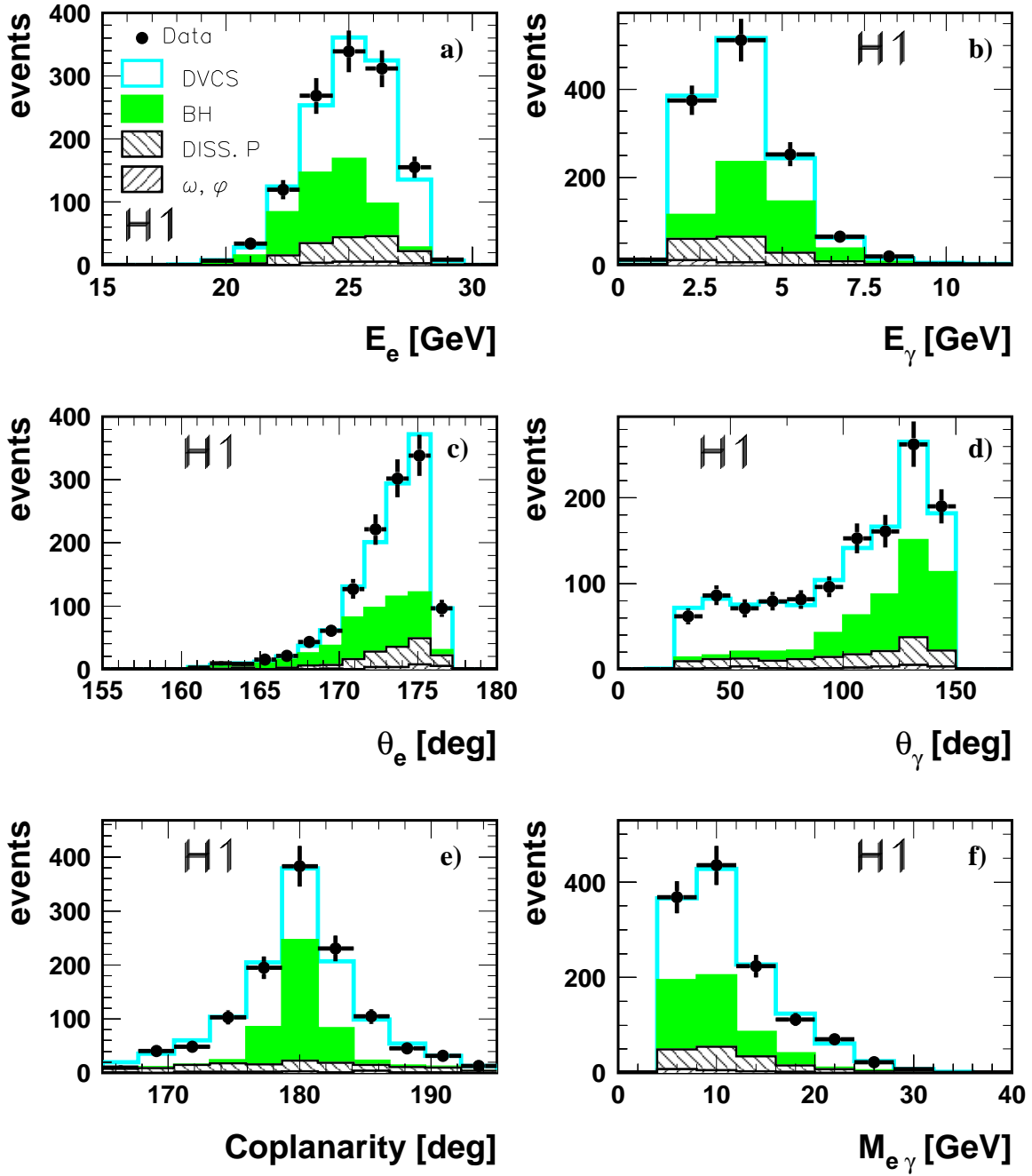,width=0.95\textwidth}
 \end{center}
 \begin{picture}(0,100)(-36,-25)
\put(33,247.0){\bf a)}
\put(112,247.0){\bf b)}
\put(34,187.0){\bf c)}
\put(112,187.0){\bf d)}
\put(33,126.0){\bf e)}
\put(112,126.0){\bf f)}
\end{picture}
 \vspace*{-10.5cm}
 \caption{\sl
Distributions of the energy of the scattered
positron (a),  the energy of the photon (b), the polar angle 
of the scattered positron (c), the polar angle 
of the photon (d), the coplanarity (e) and the positron-photon invariant
mass (f).  The data are compared with MC expectations  for
elastic DVCS, elastic BH, BH and DVCS with proton dissociation,
and \om\ and \ph\ diffractive backgrounds.
The DVCS contribution is normalised such
that the sum of the DVCS, BH and diffractive vector meson contributions
is equal to the total number of events in the data.
The normalisation of the other contributions is described in the text.
}
 \label{fig:cont}
\end{figure}

In figure~\ref{fig:cont}  the data are compared with the sum of the MC expectations.
The BH contributions and the \om\ and \ph\
backgrounds are normalised to the luminosity.
The DVCS contribution is normalised such
that the sum of the DVCS, BH and diffractive vector meson contributions 
is equal to the total number of events in the data.
The distributions of the energy and polar angle of the positron and the photon 
are shown in figures~\ref{fig:cont}a-d.
The coplanarity, shown in figure~\ref{fig:cont}e, is defined to be the 
difference of the azimuthal angles of the electron and photon directions. 
It is related to the $p_t$-balance of the positron-photon system. 
The distribution of the invariant mass of the positron and the photon
is presented in figure~\ref{fig:cont}f.
The sum of the MC contributions gives a good description of the shapes of the data 
distributions. 

\subsection{Cross Section Measurement Method} \label{sect:cross}

To extract the cross section, the selected data 
are corrected for detector efficiencies, acceptance, bin-to-bin 
migrations and for initial state radiation from the 
positron using the Monte Carlo simulation.
The inelastic BH contribution is subtracted bin by bin using the
COMPTON Monte Carlo program. The contribution of DVCS events with proton 
dissociation is subtracted bin by bin using the MILOU Monte Carlo simulation.
A 5\% correction is applied to correct for the loss of elastic DVCS events
due to the requirement that there be no signal in the forward detectors. 
The background contributions from diffractive $\omega$ and $\phi$ 
production are also subtracted using the MC simulations. 

In the leading twist approximation, the main contribution resulting from the 
interference of the BH and DVCS processes is proportional to the cosine 
of the azimuthal angle of the photon\footnote{The azimuthal angle of the photon is 
defined as the angle between the plane formed by the incoming and 
scattered positron and that formed by the $\gamma^*$ and the scattered proton.}.
Since the present measurement is integrated over this angle, 
the overall contribution of the interference term is negligible.
The elastic BH cross section can therefore be subtracted from
the total $e^+ p \rightarrow e^+\gamma p$ cross section in order to obtain 
the contribution from DVCS processes.
This contribution is then converted  to the $\gamma^* p \rightarrow \gamma p$ 
cross section using the equivalent photon approximation\footnote{After integrating over azimuthal 
angles only transversely polarised $\gamma^*$ contribute to the DVCS process.}: 

\begin{eqnarray}
 \frac { {\rm d}^3 \sigma [e p \rightarrow e \gamma p]}
       { {\rm d} y \  {\rm d} \qsq \ {\rm d} t} \  (\qsq, y,t) =
 \Gamma \, (\qsq, y) \ \ 
  \frac { {\rm d} \sigma [\gamma^* p \rightarrow \gamma p]}
       { {\rm d} t} \ (\qsq, y, t),
                                              \label{eq:gamma*p}
\end{eqnarray}
where the transverse photon flux $\Gamma$ is given by~\cite{Hand},
\begin{equation}
 \Gamma = \frac {\alpha \ (1 - y + \frac{y^2}{2}) }
     {\pi \ y \ \qsq} \qquad {\rm{with}} \qquad y = \frac {W^2+Q^2}{s} \; .
                        \label{eq:gamma}
\end{equation}
Here, $s$ is the square of the $ep$ centre-of-mass energy. 

The $t$ dependence is factorised according to:

\begin{equation}
\frac { {\rm d} \sigma [\gamma^* p \rightarrow \gamma p]}
       { {\rm d} t} \ (\qsq, y, t)
=
\frac { {\rm d} \sigma [\gamma^* p \rightarrow \gamma p] } { {\rm d} t} 
\lfrestriction{t=0} \  e^{-b|t|}\; .
                                               \label{eq:gamma*p2}
\end{equation}

The cross section $\sigma [\gamma^* p \rightarrow \gamma p]$ is extracted from
equations~\ref{eq:gamma*p} and \ref{eq:gamma*p2} using an iterative 
procedure and fitting the $t$ integrated cross section with the form:

\begin{equation}
\sigma [\gamma^* p \rightarrow \gamma p] \left(Q^2,y\right) =
N\cdot y^{\delta/2}\cdot\left(\frac{1}{Q^2}\right)^n \, ,
 \label{eq:sigfit}
\end{equation}
where $\delta$, $n$ and $b$ are free parameters and $N$ is fixed by the
integration of equation~\ref{eq:gamma*p}. More details can be found in~\cite{Rainer}. 

The same method is used to extract $\sigma [\gamma^* p \rightarrow \gamma p]$ as a
function of \qsq\ and of $x_{Bj}$.

\subsection{Systematic Errors} \label{sect:syst}
%
The main sources of systematic errors and their resulting uncertainty on
the DVCS cross section measurements are:

\begin{itemize}

\item[$\bullet$]
the subtraction of the DVCS proton dissociation background
(typically 11\% in 1996-1997, 8\% in 1999-2000 and up to 20\% in the highest $|t|$ bin) 
estimated using MC simulations with
 $b_{pdiss}~=~1.5~\pm~0.5$~GeV$^{-2}$
and an $M_Y$ dependence 
$d\sigma/dM_Y^2~\sim~(1/M_Y)^{2.0\pm0.3}$\,;
\item[$\bullet$]
the uncertainty on the acceptance correction factors (typically $10\%$ and up to $25\%$ in the 
highest $|t|$ bin) calculated by varying $b$ between $4$ and $7$~GeV$^{-2}$;
\item[$\bullet$]
the uncertainty on the determination of $\delta$ and $n$ used for the 
bin centre corrections (which ranges between $9$ and $16\%$);  
\item[$\bullet$]
the uncertainty on the BH subtraction (up to $7\%$ for the highest $W$ bin);
\item[$\bullet$]
the uncertainties on the vertex position and the measurement of the scattered positron/photon angles 
(each contribution leading to up to $12\%$ in the highest $|t|$ bin);
\item[$\bullet$]
the uncertainties on the positron/photon energies (each contribution leading to up to $12\%$ in the highest $|t|$ bin);
\item[$\bullet$]
the noise in the CJC (typically $4\%$), and in the FMD (up to 2\%);
\item[$\bullet$]
the luminosity measurement (typically $2.5\%$).

\end{itemize}

\noindent The total systematic error is found to be typically $25\%$.

\section{Results} \label{sect:results}

\subsection{Cross Sections}

The cross sections are determined separately for the two data taking periods, 
which cover different ranges in $Q^2$, and are then combined.
The 1996-1997 period covers the kinematic range $2~<~Q^2~<~20$~GeV$^2$ and
$30~<~W~<~120\,{\rm GeV}$, the 1999-2000 period $4~<~Q^2~<~80$~GeV$^2$
 and $30~<~W~<~140\,{\rm GeV}$; in both cases
$|t|~<~1\,{\rm GeV}^2$. 

The $\gamma^* p$ cross section is shown differentially in $t$ in
figure~\ref{fig:dsigdt} and given in table~\ref{tab:dsigdt} for 
$Q^2~=~4$~GeV$^2$ and $W~=~71$~GeV (using the 1996-1997 data) and 
$Q^2~=~8$~GeV$^2$ and  $W~=~82$~GeV (using the 1999-2000 data). 
The data points are fitted with 
the exponential form $e^{-b|t|}$, which gives $b~=~6.66~\pm~0.54~\pm~0.43$~GeV$^{-2}$ at
$Q^2~=~4$~GeV$^2$ where the first error is statistical and the second systematic.
At $Q^2~=~8$~GeV$^2$, a value of $b = 5.82 \pm 0.59 \pm 0.50$~GeV$^{-2}$ is obtained.
The two cross sections are averaged after correcting the 1996-1997 results to
$Q^2~=~8$~GeV$^2$ and $W~=~82$~GeV using equation~\ref{eq:sigfit} (see table~\ref{tab:dsigdt}).
The $t$ slope is then measured to be $b~=~6.02~\pm~0.35~\pm~0.39$~GeV$^{-2}$.

The cross section as a function of $Q^2$ is shown in 
figure~\ref{fig:sigq2} and given in table~\ref{tab:sigq2} for $W = 82$~GeV and $|t|<1$~GeV$^2$. 
Fitting the $Q^2$ dependence with the form $(1/Q^2)^n$ gives $n = 1.54 \pm 0.09 \pm 0.04$.
The \qsq\ dependence of the cross section is also given for a fixed value
of $x_{Bj}~=~1.8\cdot10^{-3}$ in table~\ref{tab:sigq2}, in the restricted \qsq\ range accessible for fixed $x_{Bj}$.   

The cross section as a function of $W$ is shown in
figure~\ref{fig:sigw} and given in table~\ref{tab:sigw}  
for $Q^2 = 4$~GeV$^2$ and $Q^2~=~8$~GeV$^2$; in both cases $|t|~<~1$~GeV$^2$. 
The data are fitted using the form $W^{\delta}$ which gives 
$\delta~=~0.69~\pm~0.32~\pm~0.17$ at $Q^2~=~4$~GeV$^2$ and $\delta~=~0.81~\pm~0.34~\pm~0.22$ 
at $Q^2~=~8$~GeV$^2$. The two measurements are combined as explained above
and the resulting cross section is given in table~\ref{tab:sigw}
at $Q^2~=~8$~GeV$^2$. Fitting the combined sample with the form $W^{\delta}$ gives
$\delta~=~0.77~\pm~0.23~\pm~0.19$.
The steep rise of the cross section with $W$ is a strong indication of the presence of
a hard scattering process, the value of $\delta$ being comparable to that measured in exclusive
$J/\psi$ production~\cite{jpsiprod1,jpsiprod2}. 

The extracted values of $b$, $\delta$ and $n$ are summarised in table~\ref{tab:param}.

\subsection{Discussion}

The cross section measurements from the combined data sample 
are shown with ZEUS measurements\footnote{The ZEUS
measurements, for $W~=~89$~GeV, have been rescaled to $W=82$ GeV
and from $Q^2~=~9.6$~GeV$^2$ to $Q^2~=~8$~GeV$^2$ 
using the parameter values $\delta~=~0.75$ and $n~=~1.54$ as quoted by ZEUS.}~\cite{zeusdvcs} 
and theoretical predictions as a function of
$Q^2$ in figure~\ref{fig:sigq2_c}a and as a function of $W$ in
figure~\ref{fig:sigw_c}a. 
All predictions are made assuming an exponential dependence on $|t|$, using the
measured value $b~=~6.02~\pm~0.52$~GeV$^{-2}$. The error represents the total uncertainty
of the $t$ slope which is  reflected in the band associated with each of the predicted curves.
The H1 and ZEUS measurements are seen to be consistent.
The NLO QCD calculations of Freund {\it et al.} use two different GPDs, based on
MRST 2001 and CTEQ6, for the diagonal distributions in the DGLAP domain. 
These two parameterisations show similar behaviour in $Q^2$ and in $W$ and
differ mainly in the normalisation, which reflects the relative size of the quark singlet and
gluon distributions for each set.
The H1 data are better described by the parameterisation based on CTEQ6, but it must be noted 
that the prediction also depends on the parameterisation of the ERBL region.

As shown in figures~7b and 8b, colour dipole models also provide a reasonable
description of the data, both in shape and in normalisation. 
The $Q^2$ dependence is better described by the Favart-Machado 
prediction when DGLAP evolution of the dipole (BGBK) is included. 
As regards the $W$ dependence, the H1 data are consistent with both the Donnachie-Dosch
and the Favart-Machado predictions, while the ZEUS measurements slightly favour the 
Donnachie-Dosch prediction. 

Introducing a $Q^2$ dependence of the $|t|$ slope ,
$b=b_0(1-0.15 \log(Q^2/2))$~GeV$^{-2}$~\cite{Freund:2001hd},
as extracted for exclusive $\rho$ meson production~\cite{h1_rho,zeus_rho_jpsi_hq} 
(with $b_0$ such that $b=6.02$~GeV$^{-2}$ at $Q^2=8$~GeV$^2$),
does not significantly change the above conclusions.

\section{Conclusion}

The DVCS process has been studied in the kinematic region 
$30~<~W~<~140$~GeV, $2 < Q^2 < 80$ GeV$^2$ and $|t|~<~1$~GeV$^2$ 
using data taken with the H1 detector in the years 1996 to 2000.
The $\gamma^* p \rightarrow \gamma p$ cross section 
has been measured as a function of $Q^2$ and as a function of $W$, 
and for the first time differentially in $t$.
The dependence of the cross section on \qsq\ is well reproduced by the shape $(1/Q^2)^n$ with 
$n~=~1.54~\pm~0.09~\pm~0.04$ at $W=82$~GeV. 
The $W$ dependence can be described by a fit of the form $W^{\delta}$ yielding
$\delta~=~0.77~\pm~0.23~\pm~0.19$ at $Q^2=8$~GeV$^2$. 
The  fall of the cross section differential in $t$ can be described by the form $e^{-b|t|}$ with
$b~=~6.02~\pm~0.35~\pm~0.39$~GeV$^{-2}$ at $Q^2=8$~GeV$^2$.
This first measurement of the $t$ dependence of DVCS constrains the normalisation of the theoretical predictions.
NLO QCD calculations give a good description of the normalisation as well as of the $Q^2$ and $W$ dependence of the
measured cross section using a parameterisation of the GPDs based on 
the CTEQ6 parton distribution functions. The calculations rely on ordinary (unskewed) parton distributions in the DGLAP
region and generate the skewedness dynamically.
Colour dipole model predictions also give a good general description 
of the data. This is particularly true for a saturation model in which 
the DGLAP equation is used to describe the evolution of the dipole. 

\section*{Acknowledgements}

We are grateful to the HERA machine group whose outstanding
efforts have made and continue to make this experiment possible. 
We thank the engineers and technicians for their work in constructing 
and now maintaining the H1 detector, our funding agencies for financial 
support, the DESY technical staff for continual assistance and the 
DESY directorate for the hospitality which they extend to the non-DESY 
members of the collaboration.
We are grateful to M. Diehl and A. Freund for valuable
discussions. We thank A. Freund and M. McDermott for providing the NLO QCD
predictions used in this analysis.

\clearpage
\newpage

\input{tables}


\include{figures}



\end{document}

%% file: h1auts.tex

A.~Aktas$^{10}$,               
V.~Andreev$^{26}$,             
T.~Anthonis$^{4}$,             
S.~Aplin$^{10}$,               
A.~Asmone$^{34}$,              
A.~Astvatsatourov$^{4}$,       
A.~Babaev$^{25}$,              
S.~Backovic$^{31}$,            
J.~B\"ahr$^{39}$,              
A.~Baghdasaryan$^{38}$
% ...

\bigskip{\it
% ...
$ ^{4}$ Inter-University Institute for High Energies ULB-VUB, Brussels;
          Universiteit Antwerpen, Antwerpen; Belgium$^{ c}$ \\
% ...
$ ^{10}$ DESY, Hamburg, Germany \\
% ...
$ ^{25}$ Institute for Theoretical and Experimental Physics,
          Moscow, Russia$^{ l}$ \\
 $ ^{26}$ Lebedev Physical Institute, Moscow, Russia$^{ e}$ \\
% ...
$ ^{31}$ Faculty of Science, University of Montenegro,
          Podgorica, Serbia and Montenegro \\
% ...
$ ^{34}$ Dipartimento di Fisica Universit\`a di Roma Tre
          and INFN Roma~3, Roma, Italy \\
% ...
$ ^{38}$ Yerevan Physics Institute, Yerevan, Armenia \\
 $ ^{39}$ DESY, Zeuthen, Germany \\
% ...
$ ^c$ Supported by FNRS-FWO-Vlaanderen, IISN-IIKW and IWT
      and  by Interuniversity
Attraction Poles Programme,
      Belgian Science Policy \\
% ...
$ ^e$ Supported by the Deutsche Forschungsgemeinschaft \\
% ...
$ ^l$ Partially Supported by Russian Foundation
      for Basic Research, grant    no. 00-15-96584 \\
}

%% file: tables.tex
\begin{table}[htbp]
\centering
\begin{tabular}{|c|lcc|lcc|lcc|}
 \cline{2-10}
 \multicolumn{1}{c|}{~} &\multicolumn{9}{c|}{~} \\ [-10pt]
  \multicolumn{1}{l|}{}
 & \multicolumn{9}{c|}{$d\sigma(\gamma^* p\rightarrow \gamma p)/dt
    \; \; \left[{\rm nb/GeV}^2\right]$} \\ [3pt]
 \cline{2-10}
 \multicolumn{1}{c|}{~} &\multicolumn{3}{c|}{~}  
   &\multicolumn{3}{c|}{~}  
   &\multicolumn{3}{c|}{~} \\ [-10pt]
   \multicolumn{1}{l|}{} 
 & \multicolumn{3}{c|}{1996-1997} 
 & \multicolumn{3}{c|}{1999-2000} 
 & \multicolumn{3}{c|}{All data}\\ [3pt]
 \hline
 \multicolumn{1}{|c|}{~} &\multicolumn{3}{c|}{~}  
   &\multicolumn{3}{c|}{~}  
   &\multicolumn{3}{c|}{~} \\ [-10pt]
  &
  \multicolumn{3}{c|}{$ Q^2= 4 \;$GeV$^2$} &  
  \multicolumn{3}{c|}{$ Q^2= 8 \;$GeV$^2$} &
  \multicolumn{3}{c|}{$ Q^2= 8 \;$GeV$^2$} \\  [3 pt] 
   $|t|$ $\left[{\rm GeV}^2\right] $ &
  \multicolumn{3}{c|}{$W = 71 \;$GeV} &
  \multicolumn{3}{c|}{$W = 82 \;$GeV} &
  \multicolumn{3}{c|}{$W = 82 \;$GeV} \\  [3 pt]
 \hline 
 0.1 &$\; 29.9 $ & $\;\pm 4.1 $& $\;\pm 7.1 $ &$\; 13.3 $ & $\;\pm 1.9 $& $\;\pm 3.4 $
     &$\; 12.0 $ & $\;\pm 1.2 $& $\;\pm 2.9 $\\
 0.3 &$\; 8.0 $ & $\;\pm 1.4 $& $\;\pm 1.4 $ &$\; 3.99 $ & $\;\pm 0.57 $& $\;\pm 0.69 $
     &$\; 3.44 $ & $\;\pm 0.38 $& $\;\pm 0.61 $\\
 0.5 &$\; 2.13 $ & $\;\pm 0.60 $& $\;\pm 0.69 $ &$\; 0.90 $ & $\;\pm 0.25 $& $\;\pm 0.30 $
     &$\; 0.84 $ & $\;\pm 0.17 $& $\;\pm 0.29 $\\
 0.8 &$\; 0.27 $ & $\;\pm 0.12 $& $\;\pm 0.14 $ &$\; 0.36 $ & $\;\pm 0.09 $& $\;\pm 0.14 $
     &$\; 0.21 $ & $\;\pm 0.04 $& $\;\pm 0.09 $\\
 \hline 
\end{tabular}
\caption{\sl Cross sections differential in $t$
 for the two data samples and for the combined sample.
The first errors are statistical, the second systematic.}
\label{tab:dsigdt}
\end{table}

\begin{table}[htbp]
\centering
\begin{tabular}{|c|ccc|ccc|}
 \cline{2-7}
 \multicolumn{1}{c|}{~} &\multicolumn{6}{c|}{~} \\ [-10pt]
  \multicolumn{1}{l|}{}
 & \multicolumn{6}{c|}{$\sigma (\gamma^* p\rightarrow \gamma p) \; \;
\left[{\rm nb}\right]$} \\ [3pt]
 \hline 
 ~ & \multicolumn{3}{c|}{~} & \multicolumn{3}{c|}{~}\\ [-10.0pt]
 $Q^2 \left[{\rm GeV}^2\right]$ 
  & \multicolumn{3}{c|}{$W=82$ GeV}
  &  \multicolumn{3}{c|}{ $x_{Bj}=1.8\cdot10^{-3}$}\\ [3pt]
 \hline 
 3.0  &$\; 15.7 $& $\;\pm  2.5$& $\;\pm 3.4 $ &  & & \\
 5.25 &$\; 5.7 $ & $\;\pm 1.1 $& $\;\pm 1.4 $ & $\; 6.74$ & $\;\pm 0.93$ & $\;\pm 1.02$\\
 8.75 &$\; 3.20 $ & $\;\pm 0.49 $& $\;\pm 0.69 $ & $\; 3.25$ & $\;\pm 0.51$ & $\;\pm 0.60$\\
 15.5 &$\; 1.20 $ & $\;\pm 0.22 $& $\;\pm 0.32 $ & $\; 1.45$ & $\;\pm 0.30$ & $\;\pm 0.36$\\
 25.0 &$\; 0.70 $ & $\;\pm 0.19 $& $\;\pm 0.19 $ & & & \\
 55.0 &$\; 0.15 $ & $\;\pm 0.05 $& $\;\pm 0.05 $ & & & \\
 \hline
\end{tabular}
\caption{\sl  The $\gamma^* p\rightarrow \gamma p $ cross section as a
function of $Q^2$ for $|t|<1$ GeV$^2$, at $W=82$ GeV (second column) and at
$x_{Bj}=1.8\cdot10^{-3}$ (third column).
The first errors are statistical, the second systematic.}
\label{tab:sigq2}
\end{table}

\begin{table}[htbp]
\centering
\begin{tabular}{|c|lcc|lcc|lcc|}
 \cline{2-10}
 \multicolumn{1}{c|}{~} &\multicolumn{9}{c|}{~} \\ [-10pt]
 \multicolumn{1}{l|}{} & \multicolumn{9}{c|}{$\sigma (\gamma^* p\rightarrow \gamma p)
    \; \; \left[{\rm nb}\right]$} \\ [3.0pt]
 \cline{2-10}
 \multicolumn{1}{c|}{~} &\multicolumn{3}{c|}{~}  
   &\multicolumn{3}{c|}{~}  
   &\multicolumn{3}{c|}{~} \\ [-10pt]
   \multicolumn{1}{l|}{} 
 & \multicolumn{3}{c|}{1996-1997} 
 & \multicolumn{3}{c|}{1999-2000} 
 & \multicolumn{3}{c|}{All data}\\ [3pt]
 \hline
 \multicolumn{1}{|c|}{~} &\multicolumn{3}{|c|}{~}  
   &\multicolumn{3}{c|}{~}  
   &\multicolumn{3}{c|}{~} \\ [-10pt]
 $W$ $\left[{\rm GeV}\right] $  &
  \multicolumn{3}{c|}{$Q^2 = 4 \;$GeV$^2$} &  
  \multicolumn{3}{c|}{$Q^2 = 8 \;$GeV$^2$} & 
  \multicolumn{3}{c|}{$Q^2 = 8 \;$GeV$^2$} \\ [3.0pt]
 \hline 
 45 &$\; 6.5 $ & $\;\pm 0.8 $& $\;\pm 1.1 $ &$\; 2.56 $ & $\;\pm 0.36 $& $\;\pm 0.32 $
  &$\; 2.28 $ & $\;\pm 0.21 $& $\;\pm 0.34 $\\
 70 &$\; 8.9 $ & $\;\pm 1.3 $& $\;\pm 1.6 $ &$\; 2.93 $ & $\;\pm 0.63 $& $\;\pm 0.46 $
 &$\; 2.91 $ & $\;\pm 0.35 $& $\;\pm 0.51 $\\
 90 &$\; 11.1 $ & $\;\pm 2.2 $& $\;\pm 2.7 $ &$\; 4.45 $ & $\;\pm 0.83 $& $\;\pm 0.82 $
&$\; 3.97 $ & $\;\pm 0.54 $& $\;\pm 0.85 $\\
 110 &$\; 10.1 $ & $\;\pm 4.7 $& $\;\pm 4.6 $ &$\; 5.3 $ & $\;\pm 1.4 $& $\;\pm 1.4 $
 &$\; 4.4 $ & $\;\pm 1.0 $& $\;\pm 1.5 $\\
 130 & ~ & ~ & ~ &$\; 6.4 $ & $\;\pm 2.5 $& $\;\pm 2.7 $
  &$\; 6.4 $ & $\;\pm 2.5 $& $\;\pm 2.7 $\\
 \hline 
\end{tabular}
\caption{\sl  The $\gamma^* p\rightarrow \gamma p $ cross section as a
function of $W$ for $|t|<1$ GeV$^2$ for the two data samples and for the
combined sample.
The first errors are statistical, the second systematic.}
\label{tab:sigw}
\end{table}

\begin{table}
 \begin{center}
  \begin{tabular}{|c|c|c|c|}
   \hline 
  & & & \\[-2.2ex]
 $ Q^2 $   &   $b$ [GeV$^{-2}$]  & $\delta$       & $n$ \\[0.5ex]
 \hline
 4 GeV$^2$ &  
 6.66 $\pm$ 0.54 $\pm$ 0.43 &
 0.69 $\pm$ 0.32 $\pm$ 0.17 & \\[0.2ex]
 \cline{1-3}
 8 GeV$^2$ &   
 5.82 $\pm$ 0.59 $\pm$ 0.50 &
 0.81 $\pm$ 0.34 $\pm$ 0.22 & 
 1.54 $\pm$ 0.09 $\pm$ 0.04 \\[0.2ex]
 \cline{1-3}
All data, 8 GeV$^2$ &
 6.02 $\pm$ 0.35 $\pm$ 0.39 &
 0.77 $\pm$ 0.23 $\pm$ 0.19 &
 \\[0.2ex]
 \hline
 \end{tabular}
 \end{center}
 \caption{\sl Summary of the $b$, $\delta$ and $n$ values separately for 
the two data taking periods at
$Q^2=4$ GeV$^2$ and $Q^2=8$ GeV$^2$ and for the combined sample at $Q^2=8$
GeV$^2$. 
The first errors are statistical, the second systematic.
The values of $b$ are measured
at $W=71$ GeV for $Q^2=4$ and $W=82$ GeV for $Q^2=8$. 
The values of $\delta$ and $n$ are given for $|t|<1$ GeV$^2$. 
The value of $n$ is calculated at $W=82$ GeV.}
 \label{tab:param}
\end{table}

%% file: figures.tex
\begin{figure}
 \begin{center}
  \epsfig{figure=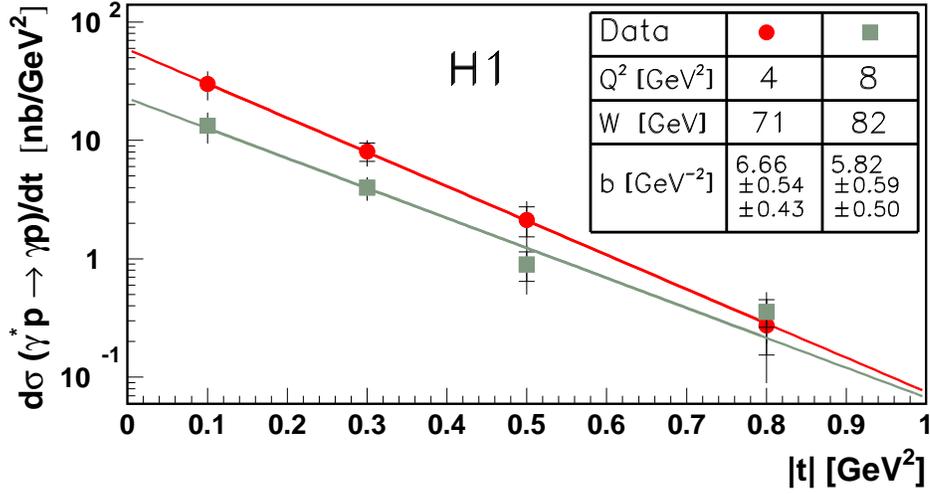,width=0.78\textwidth}
 \end{center}
 \vspace*{-0.5cm}
 \caption{\sl  The cross section $\gamma^* p \rightarrow \gamma p$ differential in $t$,
  for $Q^2=4$~GeV$^2$ at $W=71$~GeV and $Q^2=8$~GeV$^2$ at $W=82$~GeV.
  The inner error bars represent the statistical and the 
  full error bars the quadratic sum of the statistical and systematic uncertainties.
    The lines represent the results of fits to the exponential form
  $e^{-b|t|}$, giving the values of $b$ shown in the insert (see table~\ref{tab:param}).
}
 \label{fig:dsigdt}
\end{figure}

\begin{figure}
 \begin{center}
  \epsfig{figure=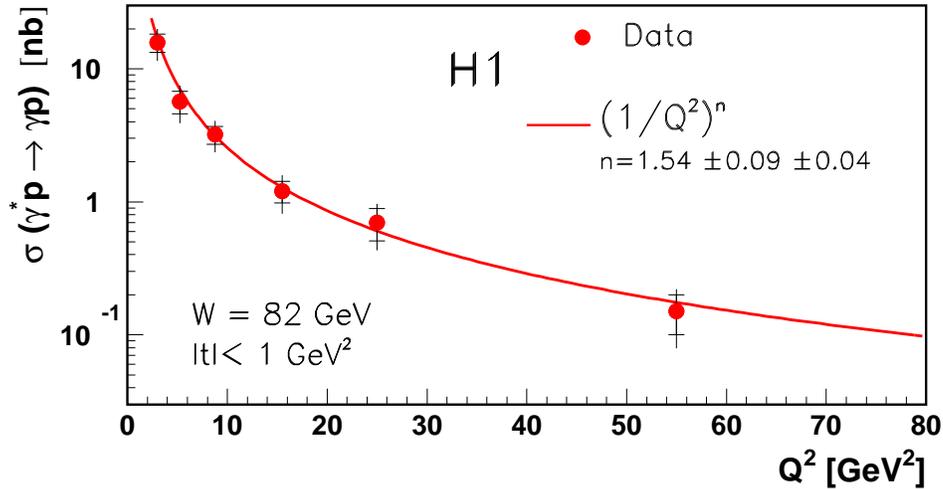,width=0.78\textwidth}
 \end{center}
 \vspace*{-0.5cm}
 \caption{\sl The $\gamma^* p \rightarrow  \gamma p$ cross section 
  as a function of $Q^2$ for $W=82$ GeV and  $|t|<1$~GeV$^2$. 
  The inner error bars represent the statistical  and the 
  full error bars the quadratic sum of the statistical and systematic uncertainties.
  The curve is the result of a fit to the form
  $(1/Q^2)^n$, giving the value of $n$ shown in the figure (see table~\ref{tab:param}).  }
 \label{fig:sigq2}
\end{figure}

\begin{figure}
 \begin{center}
  \epsfig{figure=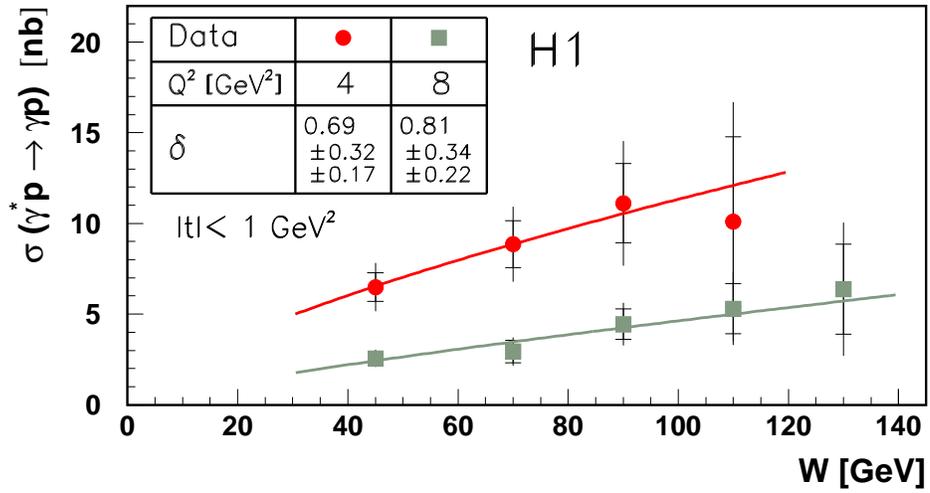,width=0.78\textwidth}
 \end{center}
 \vspace*{-0.5cm}
 \caption{\sl The $\gamma^* p \rightarrow  \gamma p$ cross section 
  as a function of $W$ for $|t|<1$~GeV$^2$ at $Q^2=4$~GeV$^2$ and
  at $Q^2=8$~GeV$^2$. 
  The inner error bars represent the statistical and the 
  full error bars the statistical and systematic uncertainties added 
  in quadrature.
  The lines are the results of a fit to the form
  $W^{\delta}$, giving the values of $\delta$ shown in the insert
  (see table~\ref{tab:param}).
  }
 \label{fig:sigw}
\end{figure}

\begin{figure}
 \begin{center}
  \epsfig{figure=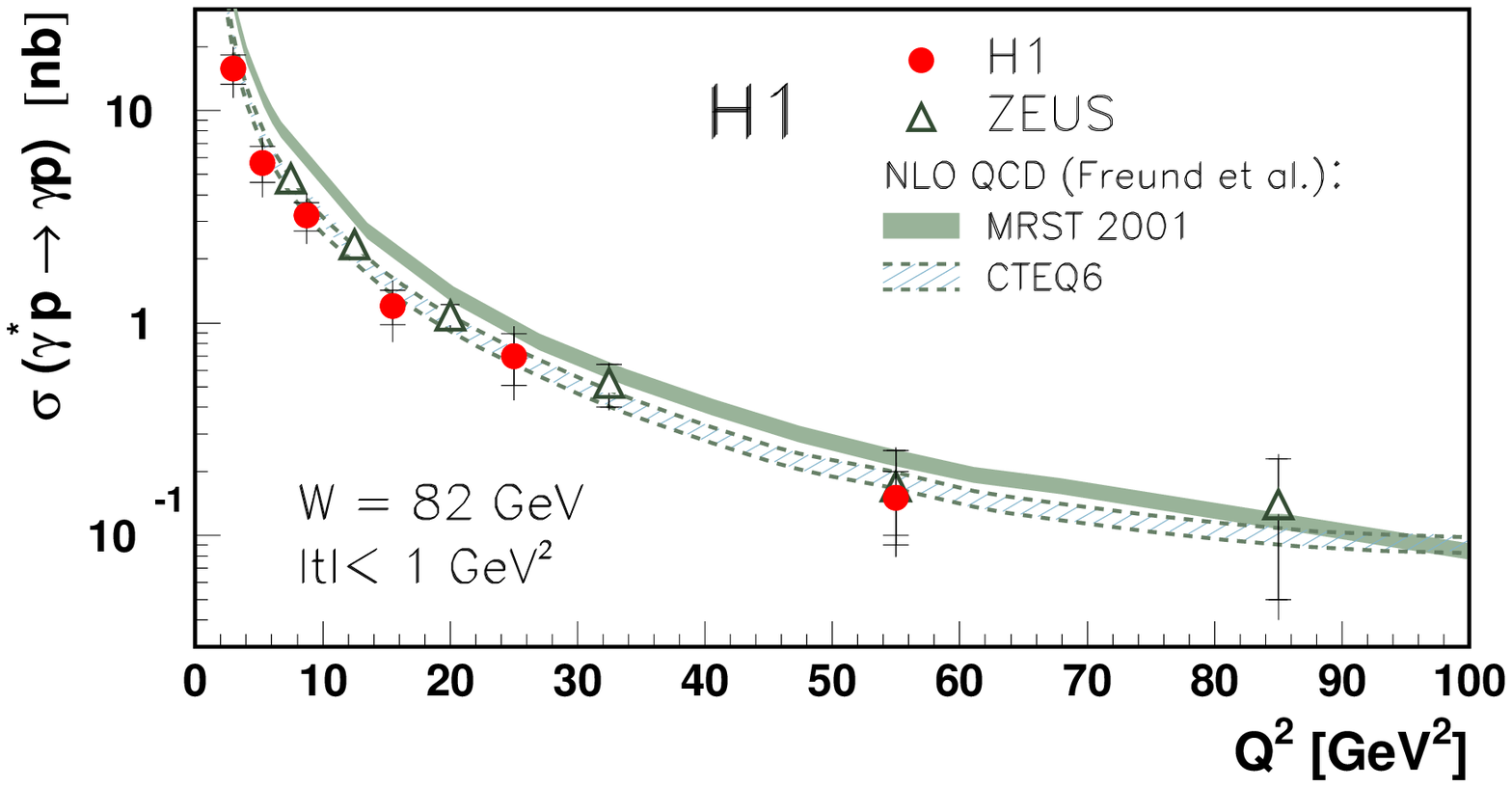,width=0.78\textwidth}\\
  \epsfig{figure=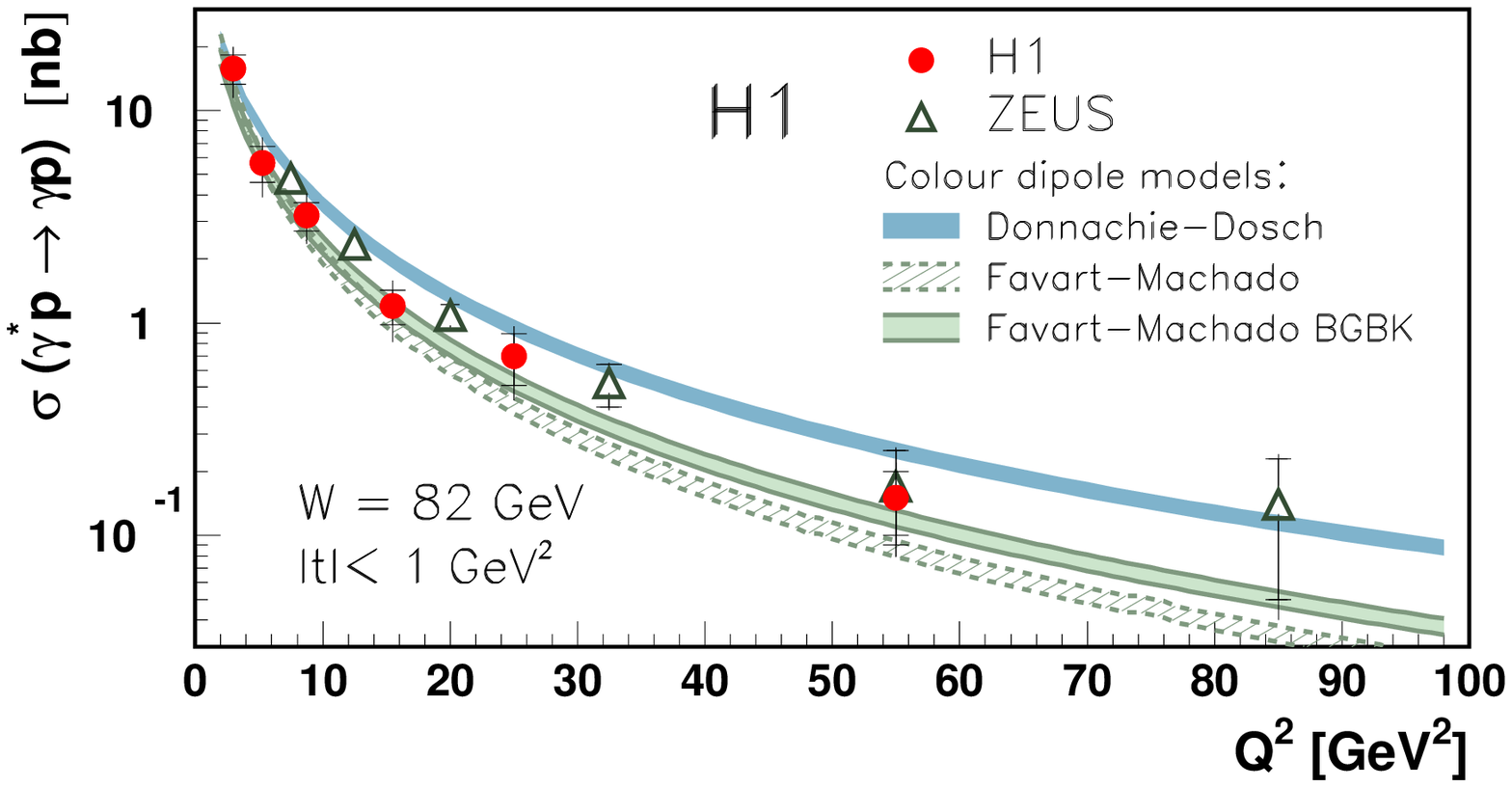,width=0.78\textwidth}\\
  \begin{picture}(0,0)(100,0)
  \put(154,135){\bf a)}
  \put(154,66){\bf b)}
  \end{picture}

 \end{center}
 \vspace*{-0.5cm}
 \caption{\sl The $\gamma^* p \rightarrow  \gamma p$ cross section 
  as a function of $Q^2$ for $W=82$ GeV and  $|t|<1$~GeV$^2$. 
  The inner error bars represent the statistical and the 
  full error bars the statistical and systematic uncertainties added 
  in quadrature.
  The H1 measurement is shown together with the results of ZEUS~\cite{zeusdvcs} and 
  several theoretical predictions. 
  a)~Comparison with QCD predictions calculated at NLO by
  Freund {\it et al.}~\cite{Freund:2002qf} based on MRST 2001 and CTEQ6 PDFs.
  b)~Comparison with the colour dipole predictions of 
  Donnachie and Dosch~\cite{Donnachie:2000px}
  and Favart and Machado with~\cite{Favart:2004uv} and 
  without~\cite{Favart:2003cu} the DGLAP evolution of
  the saturating dipole (indicated as BGBK). 
  The band associated with each prediction corresponds to the
  uncertainty on the measured $t$-slope. 
  }
 \label{fig:sigq2_c}
\end{figure}

\begin{figure}
 \begin{center}
  \epsfig{figure=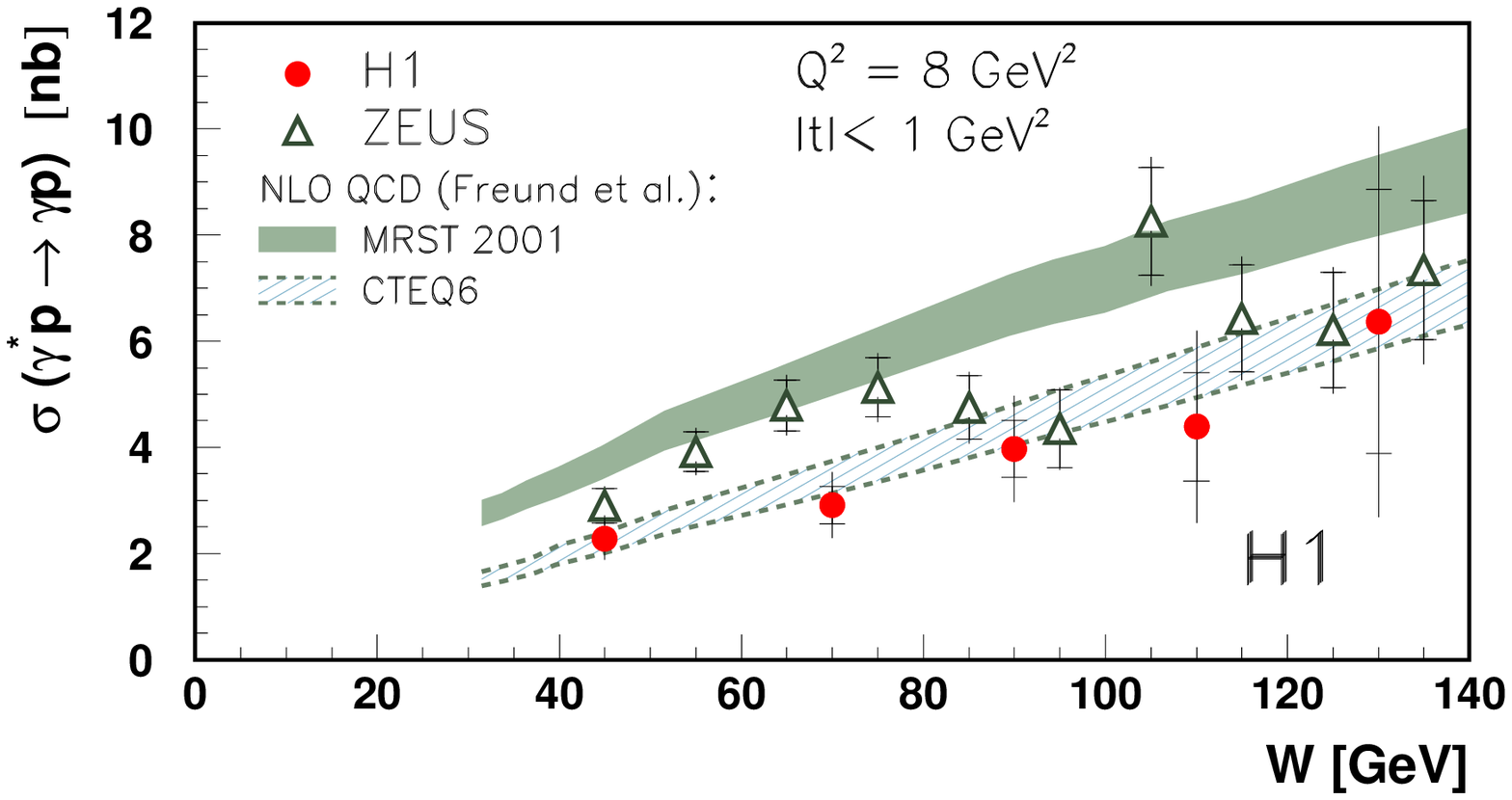,width=0.78\textwidth}
  \epsfig{figure=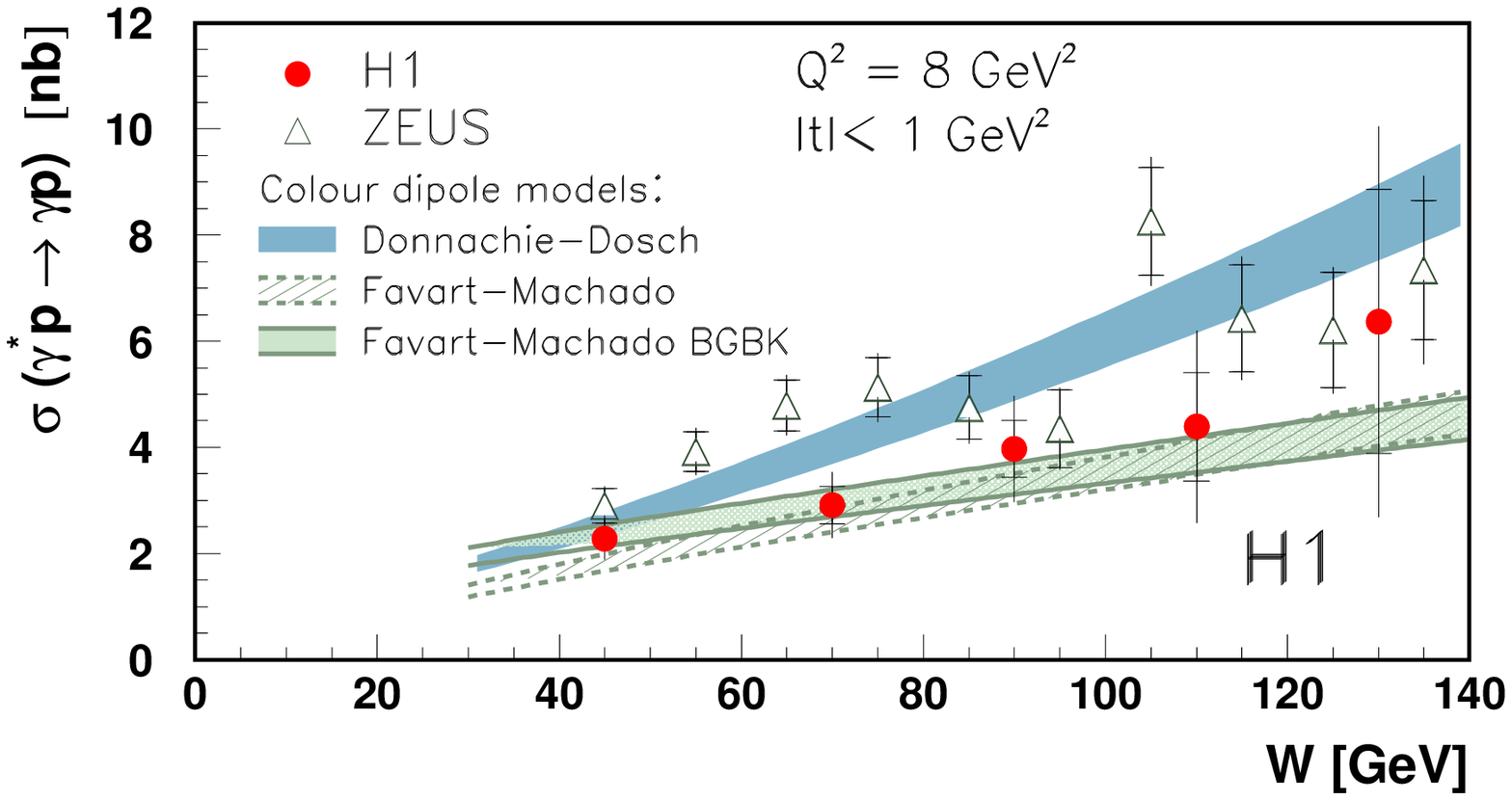,width=0.78\textwidth}\\
  \begin{picture}(0,0)(100,0)
  \put(154,135){\bf a)}
  \put(154,66){\bf b)}
  \end{picture}
 \end{center}
 \vspace*{-0.5cm}
 \caption{\sl The $\gamma^* p \rightarrow  \gamma p$ cross section 
  as a function of $W$ for $Q^2=8$~GeV$^2$ and
  $|t|<1$~GeV$^2$. 
  The inner error bars represent the statistical and the 
  full error bars the statistical and systematic uncertainties added 
  in quadrature.
  The measurement is shown with the results of ZEUS~\cite{zeusdvcs} and 
  several theoretical predictions. 
  a)~Comparison with QCD predictions calculated at NLO by
  Freund {\it et al.}~\cite{Freund:2002qf} based on MRST 2001 and CTEQ6 PDFs.
  b)~Comparison with the colour dipole predictions of 
  Donnachie and Dosch~\cite{Donnachie:2000px}
  and Favart and Machado with~\cite{Favart:2004uv} and 
  without~\cite{Favart:2003cu} the DGLAP evolution of
  the saturating dipole (indicated as BGBK). 
  The band associated with each prediction corresponds to the
  uncertainty on the measured $t$-slope. 
  }
 \label{fig:sigw_c}
\end{figure}